# Hybrid Superconductor-Quantum Point Contact Devices using InSb Nanowires

S. T. Gill[1], J. Damasco[1], D. Car[2], E.P.A.M. Bakkers[2,3,4], N. Mason[1]

1. Department of Physics, University of Illinois at Urbana-Champaign, Urbana, Illinois 61801, USA
2. Department of Applied Physics, Eindhoven University of Technology, 5600 MB Eindhoven, The Netherlands
3. QuTech, Delft University of Technology, 2600 GA Delft, The Netherlands
4. Kavli Institute of Nanoscience, Delft University of Technology, 2600 GA Delft, The Netherlands

**Abstract**

Proposals for studying topological superconductivity and Majorana bound states in nanowires proximity coupled to superconductors require that transport in the nanowire is ballistic. Previous work on hybrid nanowire-superconductor systems has shown evidence for Majorana bound states, but these experiments were also marked by disorder, which disrupts ballistic transport. In this letter, we demonstrate ballistic transport in InSb nanowires interfaced directly with superconducting Al by observing quantized conductance at zero-magnetic field. Additionally, we demonstrate that the nanowire is proximity coupled to the superconducting contacts by observing Andreev reflection. These results are important steps for robustly establishing topological superconductivity in InSb nanowires.

**Main**

    InAs and InSb nanowires (NWs) coupled to superconductors are promising material candidates for studying topological superconductivity harboring Majorana bound states[1,2] and demonstrating non-Abelian particle statistics relevant for topological quantum computation.[3] The basic procedure to observe Majorana zero modes involves tuning a superconducting proximity coupled quantum wire (i.e. a ballistic 1D system) with strong spin-orbit coupling and one spin degenerate mode in magnetic field.[1] Indeed, evidence for Majorana bound states has been observed in proximity-coupled InSb and InAs NWs as a zero-bias conduction peak in tunneling experiments.[4-7] However, preliminary experiments probing Majorana bound states in nanowires were marked by disorder and a soft superconducting gap in the tunneling regime.[4-8] Disorder in nanowire systems is known to break up ballistic transport[9,10], which is a crucial ingredient for developing 1D topological superconductivity.[1, 2, 4, 9] Additionally, disorder can produce zero-bias conductance signatures similar to Majorana bound states.[11] While the typical signature of ballistic 1D transport—quantized conductance—has been observed in InSb nanowires at high magnetic field[9] and more recently at zero-field,[19] clear demonstrations of ballistic transport at lower fields (< 1T, i.e., before the expected onset of topological superconductivity) in hybrid nanowire-superconductor systems are lacking. Hence, in order to clearly demonstrate topological superconductivity and remove alternative

mechanisms for observing zero bias conduction peaks, quantized conductance should be observed at zero magnetic field in NWs proximity coupled to superconductors.

Quantized conductance at zero-magnetic field is the step-like increase of conductance with gate voltage through a ballistic 1D constriction, in units of the quantum of conductance, $G_0 = 2e^2/h$, for each spin degenerate subband.[12] While quantized conductance in QPCs defined on 2D electron gas (2DEG) materials, such as GaAs heterostructures, is well established,[12-14] demonstrations of quantized conductance in nanowire systems are sparse. In contrast to QPCs in 2DEGs, short nanowire channels contacted by metals are more prone to backscattering effects. For example, typical processing of the nanowire-metal interface requires etching, which can lead to large structural disorder and hence a high probability of backscattering.[9-10] In addition, the mobility in nanowires can be strongly impacted by adsorbates on the uncontacted area.[15] A recent advance in nanowire synthesis has produced epitaxy of superconducting aluminum to InAs nanowires,[16-17] but residual disorder in the InAs nanowire results in unintentional quantum confined regions in these wires.[17] InSb NWs, in contrast, have higher electron mobility[9,15,18] and, as evidenced by clear demonstrations of quantized conductance,[10,19] less intrinsic disorder than InAs NWs. InSb NWs also have large Lande g factors of ~ 40-50[19], compared to values of ~5-10 in InAs;[20-22] this allows for lower magnetic fields required to induce topological superconductivity.[1-4] With optimized superconductor-nanowire interfaces, InSb NWs could be potentially a robust platform for observing and controlling 1D topological superconductivity.

In this work, we report on the development of 1D ballistic transport in InSb nanowires contacted by superconducting Al. Disorder in the nanowire is minimized by careful etching of the native oxide formed on the InSb. The low-disorder at the interface enables the hybrid superconductor-nanowire devices to behave as quantum point contacts (QPC), where we observe quantized conductance at zero-magnetic field. Additionally, we find that the level spacing of subbands in InSb nanowires differs from that seen in conventional 2DEG QPCs. In particular, we find evidence of near degeneracy in the energy spacing of the 2$^{nd}$ -3$^{rd}$ and 4$^{th}$-5$^{th}$ subbands. Finally, we demonstrate proximity effect of the superconducting contacts by observing gate-tunable Andreev reflection.

To make the devices, we use InSb nanowires (1-3 μm long, 50-80 nm diameter) grown by metal-organic phase epitaxy,[18] which are then transferred from the growth chip by use of a micromanipulator to a pre-patterned Si chip with a 300 nm silicon dioxide layer serving as a gate dielectric. The chip is cleaned using reactive ion etching prior to nanowire deposition to remove resist polymers that can degrade nanowire transport. Contacts to the nanowires are defined by electron beam lithography. Prior to metal deposition, the native oxide formed on the nanowire is removed by sulfur passivation.[23] Previous work on InSb nanowires had prepared contacts by etching the native oxide using ion milling.[4, 10, 15, 18] This procedure effectively removes the oxide to make electrical contacts, but the milling is generally harsh to the InSb crystal.[24] The homogenous etching of the InSb nanowire enabled by sulfur passivation allows deposition of thin films (<25 nm) of aluminum having a uniform morphology along the nanowire, as shown in Figure 1A. In order to prevent surface reconstruction of the Al interfacing the InSb,[16] we evaporate 5-10 nm of Ti in between a final layer of Al for our contacts. Figure 1B shows an SEM image of a completed device where the InSb nanowire is contacted by an Al/Ti/Al (20/5-10/120 nm) trilayer. Previous characterization of similarly grown InSb nanowires provided an extracted mean free path of 300 nm.[9,18] The contact spacing for the InSb nanowire/superconductor QPCs is L=150-300 nm, so the channel length is comparable to or smaller than the mean free path of the nanowire. Nanowire devices are wirebonded immediately after liftoff and left in vacuum ($\leq 10^{-2}$ mBar) for 24-48 hours to remove adsorbates before measurement in a dilution fridge.

Figure 2A (left) depicts the typical spin degenerate 1D subbands for a gate-defined QPC (*e.g.* in GaAs 2DEGs). As the Fermi energy is tuned through the subbands with a gate voltage, the resultant conductance at zero dc bias is staircase-like, as depicted in Figure 2B (left), where conductance increases with gate voltage in approximately even steps of $G_o = 2e^2/h$. In contrast, 1D constrictions with rotational symmetry, such as nanowires and nanotubes, possess a different subband structure, which results in different conductance quantization.[19,25,26,27] Rotational symmetry of the nanowire should result in a total or near degeneracy of the $2^{nd}$-$3^{rd}$ and $4^{th}$-$5^{th}$ subbands.[19,26,27] Figure 2C shows conductance vs gate voltage measurements for a nanowire QPC having a channel length of 150 nm at 10 mV applied bias. The nanowire conductance can be understood by considering the *B*=0 subband structure depicted in Figure 2A (right), which produces conductance signatures consistent with what we observe in the data (depicted in Figure 2B, right) for high bias. The data is shown for magnetic fields *B* = 0, 1, and 2 Tesla where each sweep is offset by +5 V in the plot respectively. At *B* = 0 in Figure 2C, the QPC exhibits plateaus at 1 $G_0$, 2.5 $G_0$, and 5 $G_0$. In some of the zero-field data, as in Figure 2C, a plateau at 2.5 $G_0$ is observed at finite bias as opposed to 3 $G_0$. A "half plateau" at finite bias indicates there is a difference of one subband between the source and drain electrodes. Assuming the subbands are spin-degenerate, the plateau at 2.5 $G_0$ indicates a small but finite energy separation between the $2^{nd}$ and $3^{rd}$ subbands, consistent with the subbands plotted in Figure 2A (right). As we increase magnetic field, the high bias plateau at 2.5 $G_0$ evolves to 3 $G_0$ at *B* = 2 Tesla without additional plateaus emerging. This evolution implies that degeneracy from crossings between subbands 2 and 3 occurs with increasing magnetic field. Similar plateau structure and evolution in magnetic field has recently been reported in InSb nanowires QPCs contacted by non-superconducting chromium/gold.[19] Additionally, numerical simulations also support a degeneracy from crossing between subbands 2 and 3 of InSb nanowires emerging around 2T.[19]

We observe conductance quantization consistent with a QPC having rotational symmetry in all of our InSb QPC devices measured at high bias. Figures 3A and B show the zero-magnetic field conductance quantization for two different length QPCs. Both of these devices show quantized conductance at high bias consistent with the subband spacing discussed previously Figure 3C shows the gate dependence of the conductance for the 200 nm length QPC from Figure 3A taken at 1 mV bias, which is a substantially smaller bias than that for the data shown in Figures 3A and B. In this scan, a plateau at roughly 2 $G_0$ appears between plateaus at 1 and 3 $G_0$. The imperfect quantization of the 2 $G_0$ plateau is likely related to remaining disorder in the device.[19] The plateau emerging near $2G_0$ indicates there is a finite energy difference in the spacing of the 2$^{nd}$ and 3$^{rd}$ subband that is not resolved for high bias scans.

Finally, we discuss the proximity effect of the superconducting leads to the InSb nanowire. As shown in Figure 4A, we observe a gate tunable modulation of conduction centered between $V_{SD} \approx \pm 250$ μV. Similar modulation is seen in all of our devices. For larger gate voltages (i.e., higher conductance values), the modulation develops into clearly distinguishable peaks at $V_{SD} \approx \pm 250$ μV. The peaks occur at a bias consistent with twice the induced gap commonly seen for Ti/Al leads ($\Delta_0 \approx 125$ μeV). Figure 4B shows the behavior of the differential conductance peaks near zero-bias as a function of temperature. As expected for Andreev reflection, the peaks vanish as the temperature is increased towards the superconducting transition temperature of the Al leads ($T_c \sim 1$ K). Also, the conduction peaks nearly double in magnitude below 100 mK, consistent with predictions for Andreev reflection.[28] These observations establish that the NWs are proximity coupled to superconducting leads. We note that we do not detect supercurrents across the device. Previous work on superconductivity in few mode 1D materials has consistently demonstrated small supercurrents.[29-32] The absence of clear supercurrents in our samples is likely because of an elevated electron temperature of the leads.[33] By employing a similar electronic filtering setup to previous studies of nanowire Josephson junctions,[30,32] we should be able to achieve strong correlations between conductance steps and supercurrents.

The ability to realize ballistic transport and proximity couple superconductivity in InSb NWs is an important step for robustly developing Majorana bound states in NWs. Since we observe quantized conductance plateaus at $G=G_0$, we can establish that the hybrid superconductor-InSb NW QPCs devices are tuned to the 1st mode quantum wire regime. Additionally, the superconducting contacts can be modified for high magnetic field compatibility by using a Al/NbTi/NbTiN trilayer. Hence, the procedure we have developed for making hybrid superconductor QPC devices using InSb nanowires satisfies the generic requirements for creating Majorana modes.

In conclusion, we have observed quantized conductance in InSb NWs directly interfaced with a thin layer of superconducting Al at zero magnetic field. We demonstrated that the subbands in the wire are not uniformly spaced, and that there is a near degeneracy in the dispersion relation of the second-third and fourth-fifth subbands. In addition to realizing ballistic transport in the wires, we demonstrated the proximity effect of the superconducting contacts by observing Andreev reflection. By studying similar hybrid NW-Superconductor QPCs devices at lower electron temperature, we anticipate the devices should show rich interplay between ballistic transport and superconductivity.

**Acknowledgments:** This work was supported by the Office of Naval Research grant N0014-16-1-2270. Device fabrication was done in the Materials Research Lab Central Facilities at the University of Illinois.

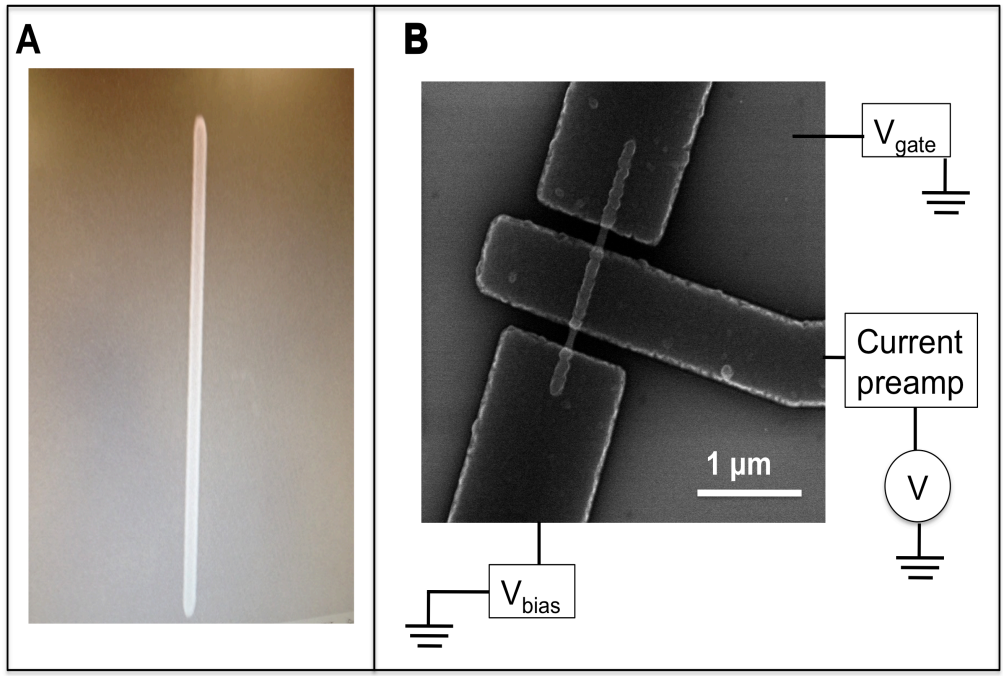

**Figure 1. A.** SEM image of a 2 μm long InSb nanowire that has been directly interfaced with 15 nm of Al. The Al shell is uniform along the entire nanowire and no grains or roughness can be observed given the resolution of the SEM. **B.** SEM image with a schematic of the measurement circuit for a completed InSb nanowire device on a Si/SiO$_2$ chip contacted by 20 nm Al/ 5 nm Ti/ 120 nm Al.

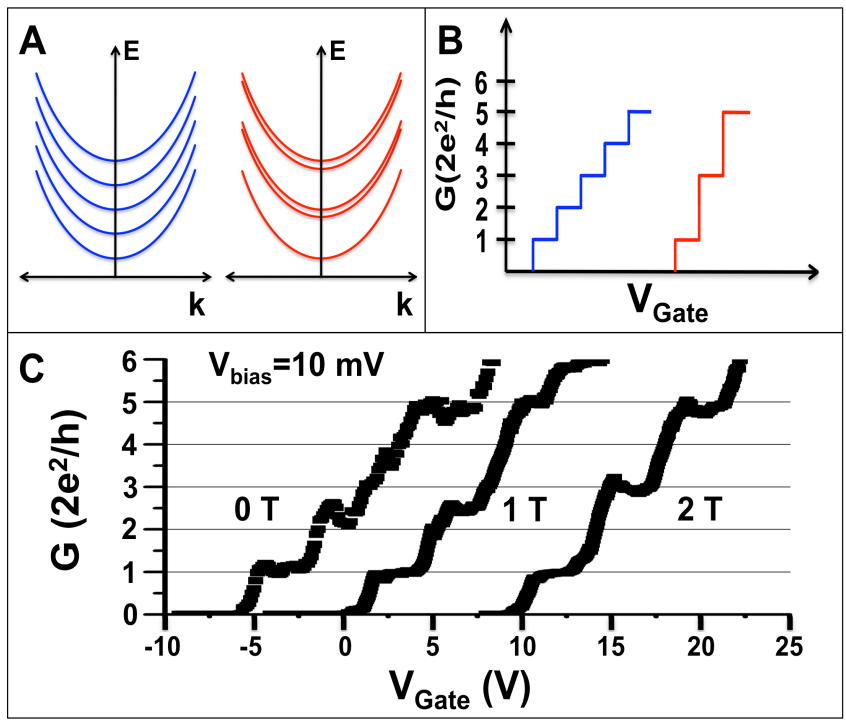

**Figure 2. A.** Diagram of the 1D subband dispersion relation for a gate defined QPC (left, in blue) and a nanowire with rotational symmetry (right, in red). **B.** Expected quantized conductance for a gate defined QPC (left, in blue) and for a rotationally symmetric QPC (right, in red) having degeneracy of the $2^{nd}$-$3^{rd}$ and $4^{th}$-$5^{th}$ subbands. **C.** Gate dependence of conductance at 10 mV bias taken at a temperature of 20 mK and for magnetic field perpendicular to the substrate. As magnetic field is increased, subband crossing creates a degeneracy of the $2^{nd}$-$3^{rd}$ subbands that produces conductance quantization identical with the conductance quantization shown on the right in **2B** for a rotationally symmetric QPC having degenerate subbands.

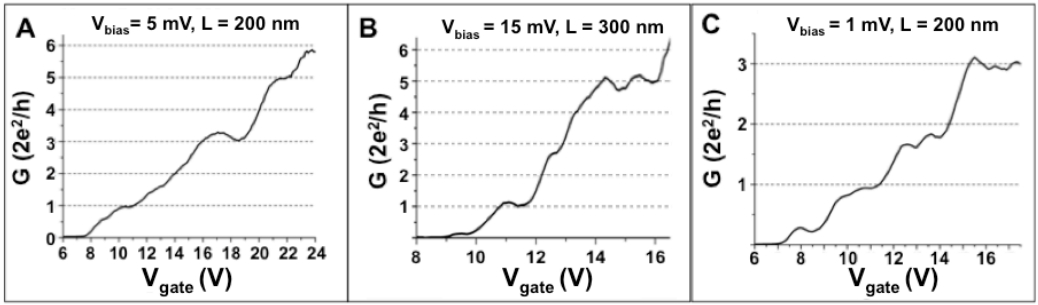

**Figure 3. A-B.** Conductance quantization at zero magnetic field for InSb QPCs having channel lengths of 200nm and 300 nm respectively. The conductance quantization is consistent with that for a QPC having rotational symmetry. **C.** At small bias ($V_{bias}$=1mV in this case), a plateau roughly at 2 $G_0$ can be resolved. The 2 $G_0$ plateau indicates that the $2^{nd}$ and $3^{rd}$ subbands are not degenerate.

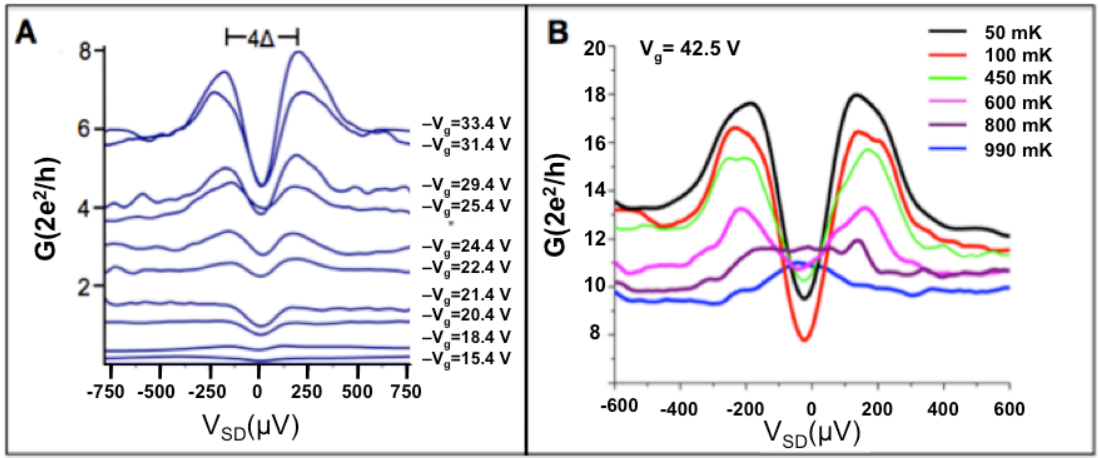

**Figure 4. A.** Differential conductance as a function of source-drain bias is plotted for different gate voltage values. Conductance is modulated around $V_{SD}$=250 µV. This enhancement occurs at a value consistent with twice the induced gap and is a signature of Andreev reflection. **B.** Temperature dependence of the conductance enhancement from Andreev reflection. As the sample temperature approaches the $T_C$ of Al, the conductance peaks thermally broaden into a single peak which nearly vanishes at $T = 990$ mK.